\documentclass[twocolumn,showpacs,preprintnumbers,amsmath,amssymb]{revtex4}

\usepackage{graphicx}
\usepackage{dcolumn}
\usepackage{bm}

\begin{document}

\title{Superconductivity in novel Ge-based skutterudites:  
$\mathbf{ \{Sr,Ba\}Pt_4Ge_{12}}$}

\author{E. Bauer$^1$, A. Grytsiv$^2$, 
Xing-Qiu Chen$^{2,4}$, N. Melnychenko-Koblyuk$^2$, 
G. Hilscher$^1$, H. Kaldarar$^1$,  
H. Michor$^1$, E. Royanian$^1$, G. Giester$^3$ M. Rotter$^2$, 
R. Podloucky$^2$ \& P. Rogl$^2$}

\affiliation{$^1$Institute of Solid State Physics, Vienna University
of Technology, A-1040 Wien, Austria}

\affiliation{$^2$Institute of Physical Chemistry, University of Vienna,
A-1090 Wien, Austria}

\affiliation{$^3$Institute of Mineralogy and Crystallography, University of Vienna,
Althanstrasse 14, A-1090 Wien, Austria}

\affiliation{$^4$School of Materials and Metallurgy, 
Northeastern University, Shenyang, 
110004, P. R. China}
			
\date{\today}

\begin{abstract}

Combining experiments and {\em ab initio} models we 
report on $\rm SrPt_4Ge_{12}$ and  $\rm BaPt_4Ge_{12}$ 
as members of a novel
class of superconducting skutterudites, 
 where Sr or Ba atoms stabilize a framework
entirely formed by Ge-atoms. 
Below $T_c=5.35$~K, and 5.10~K
for $\rm BaPt_4Ge_{12}$ and  $\rm SrPt_4Ge_{12}$, respectively,
electron-phonon coupled superconductivity  emerges, 
ascribed to intrinsic features of the Pt-Ge framework,
where Ge-$p$ states dominate the electronic structure 
at the Fermi energy.

\end{abstract}


\maketitle

Cage-forming compounds such as zeolithes, 
fullerenes, clathrates or skutterudites have been proven not only of
scientific but also of significant technological 
interest. The ability of these materials to accommodate
guest filler species constitutes a wide range 
of varying  chemical and physical properties,
 comprising magnetic order, heavy fermion
behavior, Fermi - and non-Fermi liquid 
features, or conventional or unconventional superconductivity (SC). 
For recent reviews regarding the superconducting properties 
of these classes of materials see Ref. \cite{Gunnarson,San,Imai}.
Most impressive, however,
is the exceptional thermoelectric performance 
in some of the clathrates and skutterudites \cite{Uher}. 

A structural-chemical 
description classifies
skutterudites as cage compounds. 
The simple structure, however,
acts as a prototype for a large class of 
compounds including binary as well as ternary and higher order
representatives. Hitherto, cage forming elements  
of skutterudites are essentially based on 
volatile and/or toxic pnictogens (X = P, As,
Sb). Building blocks in the framework are 
8 tilted octahedra per unit cell enclosing 
two icosahedral cages. Each
of the octahedra is centered by a transition 
metal atom like Co or its homologues (Rh, Ir).

Motivated by the known manifold of interesting 
physical properties among skutterudites, 
we searched for 
 ternary and/or higher order 
compounds exploiting the combination of (i) 
a high density of $d$-states
of a platinum group metal at the Fermi level 
with (ii) a rigid framework of rather covalently four-bonded atoms
such as Si and/or Ge. 
Thereby a novel family of Ge based skutterudites,
$\rm SrPt_4Ge_{12}$ and  $\rm BaPt_4Ge_{12}$
has been identified.
The aim of the present work is a study of the stability
and the characterization of bulk properties by means of resistivity,
magnetization, specific heat and band structure calculations. 
For both compounds, the low temperature behavior is 
dominated by the appearance of superconductivity around 5~K.

$ \rm \{Sr,Ba\}Pt_4Ge_{12}$ were prepared 
from stoichiometric amounts of high purity elements
by argon arc 
melting and subsequently heat treated in evacuated quartz
capsules at $800^{\circ} \rm C$ for two weeks. Phase purity and
lattice parameters were checked by EMPA and x-ray diffraction.
Bulk properties of these novel skutterudites
were obtained by a number of standard techniques, details are 
summarized in Ref \cite{bauer2002}.
Density functional theory (DFT) was applied using 
the Vienna {\em ab initio} simulation package 
(VASP) \cite{Kresse_1996cms,Kresse_1996prb} with 
 a fully relativistic spin-orbit 
coupling approach \cite{Blochl_1994prb,Kresse_1999prb}. 
 The Brillouin zone was sampled with 5$\times$5$\times$5
Monkhorst-Pack $\vec{k}$-point grids.
The exchange-correlation functional was treated within the
local density approximation.

The crystal structure of $ \rm \{Sr,Ba\}Pt_4Ge_{12}$ was determined 
from Kappa-CCD single crystal
X-ray data and found to be cubic,  
isotypic with the filled
skutterudite type ${\rm LaFe_4Sb_{12}}$  
\cite{Jeitschko}. 
Structure and lattice parameters are collected in Table \ref{tab1}.
Occupation factors were refined, corresponding to a full occupancy of
the Pt and Ge sublattices. 
Although not revealed from single crystal refinement, 
minor deviations (up to 3~\%) from full occupancy are possible for 
the Ba and Sr atoms. 
Since the size of the Ge-framework 
is significantly smaller than the corresponding Sb-framework, 
effective bonding between the framework cages 
(two icosahedra per unit cell) and the Ba(Sr)-center atoms is ensured. As a
consequence of this stronger bond between 
cage and guest atom, we observe very regular thermal atom displacement
factors (ADP) on all atoms. The temperature dependencies of
ADP's in the temperature region from 100 
to 300 K reveal for all atoms similar trends: typical rattling modes
caused by the heavy guest atoms in 
Sb-based skutterudites are absent in $\rm \{Ba,Sr\}Pt_4Ge_{12}$.
The structural parameters as derived from DFT calculations are in
excellent  agreement with the experimental data. 
In order to analyze the thermodynamical stability of XPt$_4$Ge$_{12}$
(X=Ba,Sr) compounds we also calculated the total energy
for a hypothetical compound Pt$_4$Ge$_{12}$. 
The bonding energy $E_\mathrm{X}$ for guest atom X
is obtained from the relation
$E_\mathrm{X} = U^{DFT}_{\mathrm{XPt}_4\mathrm{Ge}_{12}} -
U^{DFT}_{\mathrm{Pt}_4\mathrm{Ge}_{12}} -  U^{DFT}_\mathrm{X}$ 
in which  $U^{DFT}$ denotes the corresponding total energy of the compound
or elemental solid in its equilibrium state.
The values $E_{\mathrm{Ba}}$ = -3.24 eV and $E_{\mathrm{Sr}}$ = -3.38 eV
emphasize the stabilizing effect of the Ba and Sr guest atoms.

\begin{figure}[!htb]
	\centering
		\includegraphics[width=0.48\textwidth]{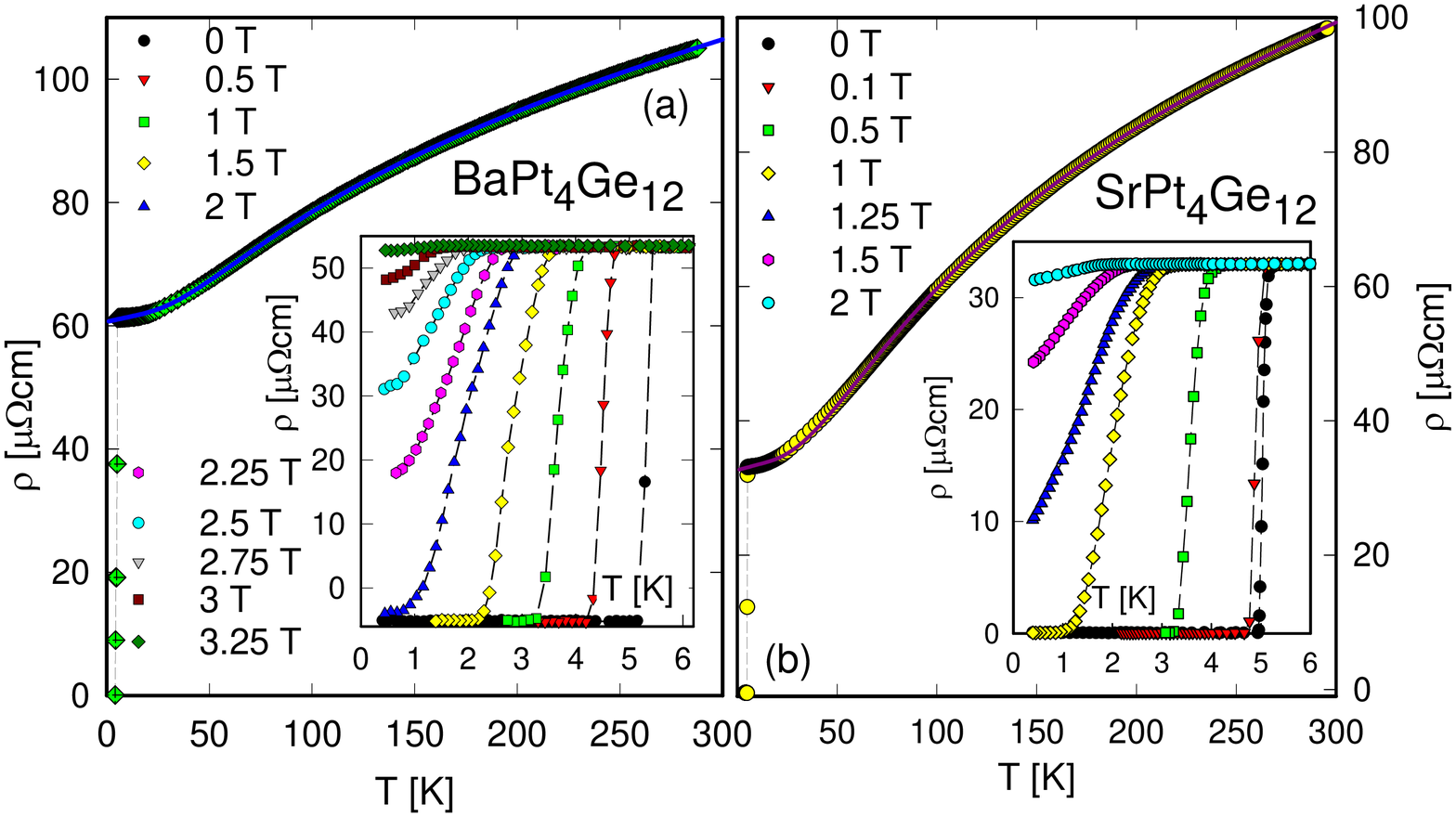}
	\caption{(Color online) Temperature dependent electrical resistivity $\rho$ of 
	$\rm BaPt_4Ge_{12}$ (a) and 	$\rm SrPt_4Ge_{12}$ (b). 	
	Both insets show the field dependence of $\rho (T)$.} 
	\label{fig2}
\end{figure}

Superconductivity is found for $\rm BaPt_4Ge_{12}$ 
and $\rm SrPt_4Ge_{12}$ from resistivity measurements 
using polycrystalline samples at $T_c \approx 5.3$~K 
and 5.1~K, respectively (compare Fig. \ref{fig2}(a,b)).
The normal state regions, $T > T_c$ of these ternary compounds do not behave
like simple  metals, since the standard model of the electrical 
resistivity of metallic systems, i.e., the Bloch-Gr\"uneisen formula 
is not applicable. Such observations were made in many superconducting 
materials and may be attributed to a substantial electron-phonon
interaction strength, responsible for the formation of Cooper pairs
in conventional superconductors. 
Rather, the overall $\rho (T)$-dependence of both skutterudites
follows the model of Woodward and Cody \cite{Cody}, 
which initially was applied to A15 superconductors such as $\rm Nb_3Sn$.
Least squares fits to this model are shown as solid lines in both figures,
revealing reasonable agreement for characteristic 
temperatures $T_0 = 123$ and 121~K, respectively. 
The differences of the residual resistivities 
may correspond to small differences in the filling of the 
$2(a)$-sites by Ba or Sr. In case of the smaller atom Sr,
filling seems to be more complete. 
The application of a magnetic field suppresses superconductivity 
of $\rm BaPt_4Ge_{12}$ at fields above 2~T,
while for  $\rm SrPt_4Ge_{12}$ the upper critical field $H_{c2}(0) \approx 1$~T
(insets of Fig. \ref{fig2}).

\begin{table}[h]
\centering
\caption{Normal state and superconducting properties of $\rm BaPt_4Ge_{12}$
	and $\rm SrPt_4Ge_{12}$ which crystallize in the cubic 
skutterudite structure: space group {Im\={3}, (No. 204)}; Ba and Sr 
are at the 2\textit{(a)} (0, 0, 0) sites, Pt at the 
8\textit{(c)} ($\frac{1}{4}$, $\frac{1}{4}$, $\frac{1}{4}$) sites, 
and Ge at the 24\textit{g} (0, y, z) sites. $U_{eq}$ is a mean value of the atomic
displacement ellipsoid.} 
\begin{tabular}{|l||c|c|} \hline 
property	& $\rm BaPt_4Ge_{12}$ & $\rm SrPt_4Ge_{12}$  \\ \hline \hline
lattice parameter $a~@300$~K~[nm]  & 0.86928(3)	&	0.86601(3)   \\	
Ge 24{\textit g} site: y & 0.15302 & 0.15197 \\
Ge 24{\textit g} site: z & 0.35683 & 0.35536 \\
$R_{F2}  = \sum \vert F_o^2 - F_c^2 \vert  / \sum F_o^2$ & 0.019 & 0.018 \\
$U_{eq}$(\{Ba,Sr\}) [nm$^2$]    &  0.000066(2)           &   0.000118(3)   \\
$U_{eq}$(Pt) [nm$^2$]          &     0.000062(1)           &   0.000069(1)   \\
$U_{eq}$(Pt) [nm$^2$]          &     0.000092(2)           &  0.000097(2)   \\
transition temperature $T_c$ [K]   &  5.35	&	5.10      \\		
upper critical field $\mu_0 H_{c2}$~[T]  &    1.8        &         1       \\
thermod. critical field $\mu_0 H_c$~[mT]      &    53        &        52      \\
Fermi velocity $v_F$~[m/s]                 &    52500     &        67000   \\
coherence length $\xi$~[nm]             &    14(1)       &        18(1)    \\
penetration depth $\lambda$~[nm]				&    320(10)        &        250(10)     \\
G.L. parameter $\kappa$                 &    24(1)      &        14(1)   \\
			\hline \hline
		\end{tabular}
	\label{tab1}
\end{table}

\begin{figure}[b]
	\centering
		\includegraphics[width=0.48\textwidth]{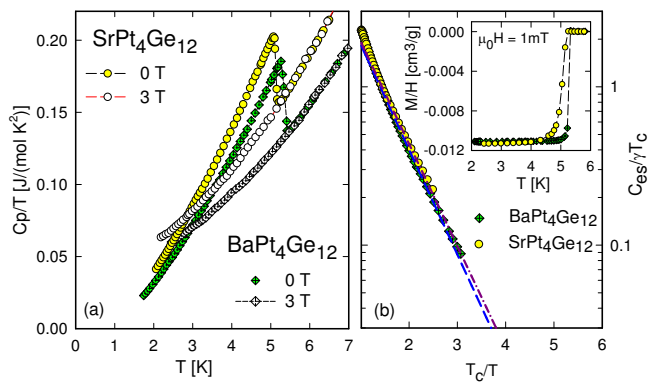}
	\caption{(Color online) (a) Temperature dependent specific heat $C_p$ of $\rm SrPt_4Ge_{12}$
	and $\rm BaPt_4Ge_{12}$ plotted as $C_p/T$ vs. $T$ for 0 and 3 T.
	(b) Semi-logarithmic plot 
  $C_{es}/\gamma T_c $ vs. $T_c /T$. The inset shows the low temperature and low field 
  susceptibility $M/H$ for both compounds.}
	\label{fig4}
\end{figure}

\begin{figure}[tb]
	\centering
		\includegraphics[width=0.48\textwidth]{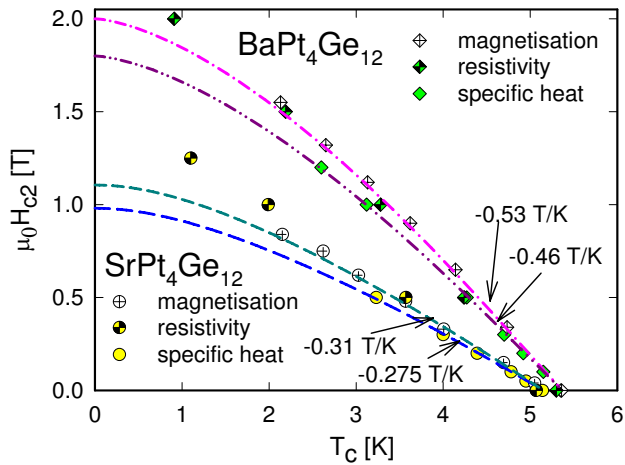}
	\caption{(Color online) (a) Temperature dependent upper critical field
	 $\mu_0 H_{c2}$ of $\rm BaPt_4Ge_{12}$ and $\rm SrPt_4Ge_{12}$ as obtained from 
	 resistivity, magnetization and specific heat measurements. The dashed and the 
	 dashed-dotted lines represent the WHH model.}
	\label{fig5}
\end{figure}

The susceptibility $\chi$
exhibits a rather sharp transition at $T = 5.3$ and 5.1~K 
[inset, Fig. \ref{fig4}(b)], 
dropping from an initially small positive value to the diamagnetic value of 
$[-1/(4 \pi)]$ for zero field cooling, which  
corresponds to a full Meissner Ochsenfeld effect. 
Magnetization measurements performed 
at various temperatures and 
magnetic fields up to  6~T evidence  type II superconductivity.
Upper critical field values are summarized in Fig. \ref{fig5}.

The heat capacity, $C_p$, of $\rm \{Sr,Ba\}Pt_4Ge_{12}$
 is plotted in Fig. \ref{fig4} as $C_p/T$ vs. $T$ for  zero and 3 T. 
For sake of clarity, measurements taken at intermediate field
values are not shown here.
The jump of $C_p(T)$ below 6~K evidences bulk  
superconductivity in both cases. An
idealization of the heat capacity anomaly under the constraint
of entropy balance between the superconducting and the normal state yields
$T_c = 5.35$ and 5.1~K for the Ba and Sr based compounds, respectively. 
Assuming that the specific heat 
of metallic compounds at low temperature follows
$C_p = \gamma T + \beta T^3$, ($\gamma$ is the Sommerfeld 
coefficient and $\beta$ is proportional to  the Debye temperature $\theta_D$),
least squares fits 
reveal $\gamma = 42$~mJ/molK$^2$ and $\theta_D = 247$~K (Ba) and 
$\gamma = 41$~mJ/molK$^2$ and $\theta_D = 220$~K  (Sr).
It is worth to be noted that the Debye temperature of  $\rm BaPt_4Ge_{12}$
is larger than that of  $\rm SrPt_4Ge_{12}$.
In general, however, materials with smaller masses exhibit larger
Debye temperatures. This anomaly observed may correspond to the fact that, 
while the volume of the unit cells of both compounds differ by only 1\%,
the atomic volumes of Sr and Ba differ by about 12~\%. This causes 
a weaker  bonding of Sr to the framework, hence a weaker force 
constant may result in lower values of $\theta_D$.

Taking into account the McMillan model \cite{McMillan} allows 
calculation of  the dimensionless electron-phonon coupling constant
$\lambda$. Setting the repulsive screened Coulomb term $\mu^* \approx 0.13$,
yields in both cases $\lambda \approx 0.7$. This refers 
to superconductors well beyond the weak coupling limit. In comparison, 
$\mu^*$ of different cage forming compounds covers a range from $\approx 0.1$
to $\approx 0.3$ \cite{Gunnarson,San,Imai}.

 The jump of the specific heat 
 $\Delta C_p/T(T = T_c) \approx 58$mJ/molK$^2$ (Ba) and $\approx 57$mJ/molK$^2$ (Sr)
allows calculation of
$\Delta C_p/(\gamma_n T_c) \approx 1.35$, which is near to
the figure expected from  BCS theory 
[$\Delta C_p/(\gamma T_c) \approx 1.43$]. 
As the magnetic field strength increases (not shown here), 
both the transition temperature
and the anomaly right at $T_c$ are suppressed,
constituting the phase diagram  in Fig. \ref{fig5}. 

The superconducting gap $\Delta (0)$ can be derived from a 
comparison of the modified BCS expression,   
$C_{es}(T)=8.5\gamma T_c\exp (-0.82\Delta (0)/k_{\rm B}T)$ 
with the experimental data 
depicted in a semi-logarithmic plot 
$C_{es}/\gamma T_c $ vs. $T_c /T$ in Fig. \ref{fig4}(b)
where $\rm BaPt_4Ge_{12}$ and $\rm SrPt_4Ge_{12}$ 
exhibit for $T_c/T>2$ an exponential temperature 
dependence indicating a  
ratio $\Delta (0)/k_{\rm B}T_c \approx 1.8$ in close 
agreement with the BCS value $\Delta_{\rm BCS} (0)=1.76k_{\rm B}T_c$. 

The thermodynamic critical field is calculated from the free energy difference 
between the superconducting and normal state:
$\Delta F (T) = F_n - F_s = \mu_0 H_c^2 (T)/2$,
where $F_n$ and $F_s$ are evaluated from the specific heat 
in the normal and superconducting state, respectively.
The values obtained  are  $\mu_0H_c(0) \approx 53\,(2)$ and 52\,(2)\,mT for 
Ba and the Sr-based compound, respectively.

Fig. \ref{fig5} displays the temperature dependent upper critical field
$\mu_0H_{c2}$ of $\rm \{Sr,Ba\}Pt_4Ge_{12}$ as deduced from 
field dependent resistivity,  magnetization and heat capacity measurements. 
The slopes of the upper critical field $\partial (\mu_0 H_{c2}) / \partial T \equiv
\mu_0 H _{c2}'$ are collected in Table \ref{tab1}, yielding slightly
larger values deduced from  magnetization 
and resistivity data than those from specific heat,
which may be attributed to pinning and surface effects.
$\mu_0 H _{c2}'$ of  $\rm BaPt_4Ge_{12}$ is 
larger than $\mu_0 H _{c2}'$ of $\rm SrPt_4Ge_{12}$.

Primarily, two mechanisms are 
responsible for the limited value of $\mu_0 H_{c2} $: orbital pair 
breaking and Pauli limiting. 
Werthamer et al. \cite{Werthamer} derived an expression 
({\em WHH model}) for the 
upper critical field $\mu_0 H_{c2}$
in terms of orbital pair-breaking, 
including the effect of Pauli spin para\-magnetism and
spin-orbit scattering. A comparison of the experimental results
with the WHH model is based on two parameters,
$\alpha$, the Pauli paramagnetic limitation ({\em Maki parameter}) and 
$\lambda_{so}$ describing spin-orbit scattering. If the 
	atomic numbers of the elements constituting the material increase,
$\lambda_{so}$ is expected to increase as well.   

The Maki parameter $\alpha$ can be estimated via 
the Sommerfeld value $\gamma$ 
and $\rho_0$ \cite{Werthamer}, i.e.,
$
\alpha = (3 e^2 \hbar \gamma \rho_0)/(2 m \pi^2 k_B^2)
$
with $e$ the electron charge and $m$ the electron mass. 
Taking the experimental $\rho_0$ and $\gamma$ 
yields $\alpha = 0.18$ for $\rm BaPt_4Ge_{12}$ and 
$\alpha = 0.14$ for $\rm SrPt_4Ge_{12}$. 
 We have adjusted the WHH model to the 
experimental data (dashed and dashed-dotted lines in Fig. \ref{fig5}),
revealing $\lambda_{so} \approx 15$ for all data-sets.

The thermodynamic and the upper 
critical field are used to calculate 
the ratio of the penetration depth  
$\lambda_{\rm GL}(0)$ to the coherence length $\xi_{\rm GL}(0)$ via 
Abrikosov's relation 
$\lambda_{\rm GL}(0)/\xi_{\rm GL}(0)\equiv\kappa_{\rm GL}(0)=
H_{c2}/[\sqrt{2}H_c(0)]$  yielding the Ginzburg-Landau parameter 
$\kappa_{\rm GL}= 24$\,(2) and 14\,(2) for the Ba and Sr-based compounds.
The absolute values of the coherence length $\xi_0$ and the penetration depth
$\lambda (0)$ can be evaluated via the isotropic
Ginzburg-Landau-Abrikosov-Gor'kov (GLAG) theory.
Values obtained are presented in Table \ref{tab1}.

The Fermi velocity $v_F$ can be calculated from the Fermi surface area $S_s$
as shown by Orlando et al. \cite{Orlando}. Values are given in Table \ref{tab1}.   
Combining  $S_s$ and $\rho_0$, 
a mean free path $l_{tr}$ of about $\approx 1.0 \cdot 10^{-8}$~m 
and $\approx 1.4 \cdot 10^{-8}$~m for the Ba and Sr compound, respectively, is derived.

The fact that $\mu_0H_{c2}'$ of $\rm BaPt_4Ge_{12}$
is larger than that of $\rm SrPt_4Ge_{12}$ can be understood 
in terms of the Ginzburg-Landau theory. Hake \cite{Hake} and 
Orlando et al. \cite{Orlando} derived a model equation for $\mu_0H_{c2}'$
which primarily depends on two parameters: on the inverse of
the square of $v_F$ and on the inverse of $l_{tr}$.  
Taking into account the parameters deduced from the above
analyses explains  in both cases, at least qualitatively, 
differences of $\mu_0H_{c2}'$. While the decrease
of $v_F$ from the Sr to the Ba case can be conceived by an increase
of the unit cell volume ($v_F \propto (N/V)^{1/3}$, for free elelctrons),
the decrease of the mean free path corresponds to the 
increase of the residual resistivity
from $\rm SrPt_4Ge_{12}$ to $\rm BaPt_4Ge_{12}$.

\begin{figure}[tb]
\begin{center}
		\includegraphics[width=0.48\textwidth]{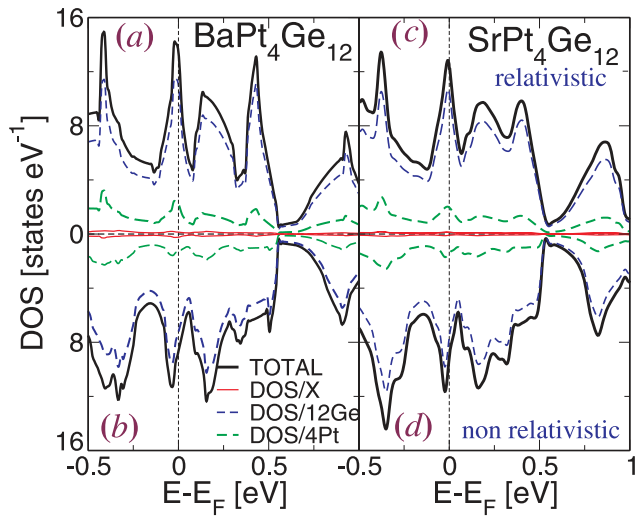}
\caption{(Color online) DOS for BaPt$_4$Ge$_{12}$ (a,b)
and SrPt$_4$Ge$_{12}$ (c,d) as derived from DFT calculations. Fermi
energy is situated at zero energy.  
The upper half (a),(c) shows the density of states
for a fully relativistic calculation, which includes spin-orbit (SO) coupling. The
lower half (b),(d) of the figure presents the DOS of a standard non-relativistic
calculation. The total DOS is decomposed into the local DOS representative for
one Ba(Sr), four Pt, and twelve Ge atoms.  
The relativistic effect is small but important, 
because it shifts the Fermi level of the non-relativistic DOS almost into the
maximum of the relativistic DOS. 
\label{fig6}}
\end{center}
\end{figure}

From $l_{tr}/\xi \approx 1$ we classify  $\rm \{Ba,Sr\}Pt_4Ge_{12}$ as 
a dirty limit superconductor, and  $\kappa$ of the order of 10 to 20 refers
to a pronounced type II superconducting behavior.

The DOS in Fig. 4 reveal  the individual contributions of X=(Ba,Sr), Pt and Ge
atoms showing that Ge states, which are of $p-$like character,
are dominating at Fermi energy.
The Ge states hybridize with Pt $5d$-like states by which the
spin-orbit coupling effect is  transmitted to the DOS at E$_F$.
Consequently, by making use of the electron localization function technique 
\cite{Silvi and Salvin}, expressive Ge-Ge covalent 
bonds and typical metallic Pt bonding are being deduced.
The metallic features of the DOS around $E_F$ 
convincingly confirm that the  Zintl concept no longer applies
to $\rm \{Sr,Ba\}Pt_4Ge_{12}$, while its applicability  is quite obvious 
among pnictogen-based skutterudites.

The total DOS at $E_F$ can be compared with the Sommerfeld 
value of the specific heat $\gamma = \frac{1}{3}\pi^2 N(E_F)k_B^2$.
The experimental values slightly larger than 40~mJ/molK$^2$ for 
the Ba and Sr compounds, 
along with an electron-phonon enhancement factor $\lambda_{ep} =0.7$ 
(estimated via the McMillan formula \cite{McMillan}) would require a 
bare DOS equivalent to  $\approx 25$~mJ/molK$^2$,
which compares favorably with the DOS calculations involving spin-orbit  
coupling: 31~mJ/molK$^2$ for the Ba ($N(E_F)$ = 13.2 states eV$^{-1}$ f.u$^{-1}$) 
and 28~mJ/molK$^2$ for the Sr compound
($N(E_F)$ = 12.1 states eV$^{-1}$ f.u$^{-1}$). Without 
spin-orbit coupling, the specific heats, $\gamma$, are 19.5 and   
21.0 ~mJ/molK$^2$ for Ba and Sr, respectively. 
This implies the importance of the 
relativistic effects.

In summary, superconducting $\rm \{Sr,Ba\}Pt_4Ge_{12}$ are the first
skutterudites where the framework in the structure is entirely
built by Ge-atoms. 
DFT calculations proved that X=(Ba,Sr) guest atoms strongly stabilize the
compounds.  Most strikingly, the calculated DOS  around E$_F$ is composed of  
hybridized Ge $4p$-like and Pt $5d$-like states, 
and it has a sharp peak with its maximum very close
to E$_F$.
The influence of the guest atoms (Ba or Sr) on
superconductivity, however, may be ruled out due to the fact that (i) the
Ba- or Sr-like DOS around $E_F$ is negligible  and (ii) the DOS around E$_F$
for the  \emph{hypothetical}  X-free Pt$_4$Ge$_{12}$ framework is very
similar to the one of  XPt$_4$Ge$_{12}$ (X = Ba and Sr).
Hence, superconductivity appears to be an intrinsic property
of the Pt-Ge cage-forming structure.
This conclusion is in line with the 
slightly smaller value of $T_c$ observed in  $\rm SrPt_4Ge_{12}$,
in marked contrast to the isotope effect, where lighter
masses would originate larger SC transition temperatures.

Work supported by the Austrian science foundation 
FWF, project No. P19165 and P16778/2
and by CMA, {\em Complex Metallic Alloys} (EU contract
NMP3-CT-2005-500140). 
X.-Q.C. is grateful to the 
National Nature Science Fund of China 
(Project No. 50604004).


\begin{thebibliography}{99}


\bibitem{Gunnarson} O. Gunnarson, Rev. Mod. Phys. {\bf 69}, 575 (1997). 

\bibitem{San} A. San-Miguel and P. Toulemonde, High Pressure Res. {\bf 25},
159 (2005).

\bibitem{Imai} M. Imai et al. Phys. Rev. B {\bf 75}, 184535 (2007).

\bibitem{Uher} C. Uher, 
Semiconductors and Semimetals  {\bf 69}, 139 (2001).


\bibitem{bauer2002} E. Bauer et al., Phys. Rev. B {\bf 66}, 214421 (2002).

\bibitem{Kresse_1996cms}
G.~Kresse and J.~Furthm\"uller,
\newblock {Comput. Mater. Sci.} {\bf 6}, 15 (1996).

\bibitem{Kresse_1996prb}
G.~Kresse and J.~Furthm\"uller,
\newblock {Phys. Rev. B} {\bf 54}, 11169 (1996).

\bibitem{Blochl_1994prb}
P.~E.~Bl\"ochl,
\newblock {Phys. Rev. B} {\bf 50}, 17953 (1994).

\bibitem{Kresse_1999prb}
G.~Kresse and D.~Joubert,
\newblock {Phys. Rev. B} {\bf 59}, 1758 (1999).


\bibitem{Jeitschko} W. Jeitschko,  D.J. Braun, 
Acta Crystallogr. B   {\bf 33}, 3401 (1977).


\bibitem{Cody} D.W. Woodward and G.D. Cody, Phys. Rev. {\bf 136}, (1964) A166. 

\bibitem{McMillan} W.L. McMillan,
Phys. Rev.  {\bf 167}, 331 (1968).


\bibitem{Werthamer} N.R. Werthamer, et al., 
Phys. Rev. {\bf 147}, 295	(1966).


\bibitem{Orlando} T.P. Orlando,  et al.,  Phys. Rev. B {\bf 19}, 4545 (1979). 


\bibitem{Hake} R.R. Hake, Phys. Rev. {\bf 158}, 356 (1967). 


\bibitem{Silvi} B. Silvi and A. Savin, Nature, {\bf 371}, 683 (1994).


\end{thebibliography}
\end{document}